\documentclass[prd,amsmath,amssymb]{revtex4}

\usepackage{graphicx}% Include figure files
\usepackage{bm}% bold math

\normalbaselines \topmargin -0.25 truein \oddsidemargin 0.30
truein \evensidemargin 0.30 truein 
\begin{document}

\title{Nonlinear interaction of the axion field with dynamic aether \\ and SU(2) symmetric gauge field in the anisotropic Universe}

\author{Alexander B. Balakin}
\email{Alexander.Balakin@kpfu.ru} \affiliation{Department of General
Relativity and Gravitation, Institute of Physics, Kazan Federal University, Kremlevskaya
str. 18, Kazan 420008, Russia}
\author{Gleb B. Kiselev}
\email{gleb@karnaval.su} \affiliation{Department of General
Relativity and Gravitation, Institute of Physics, Kazan Federal University, Kremlevskaya
str. 18, Kazan 420008, Russia}

\date{\today}

\begin{abstract}
Based on the concept of the dynamic aether emergence as a result of spontaneous polarization of the color aether, we consider the SU(2) symmetric theory of interaction of the gauge and axion fields in the framework of anisotropic cosmological model of the Bianchi-I type. We focus on the analysis of the non-Abelian analog of the U(1) symmetric model of axionically induced generation of the electric field in a magnetized medium.
\end{abstract}

\maketitle

\section{Introduction}\label{Intro}

In the recent papers \cite{1,2} we have formulated the general relativistic theory of nonlinear interaction between pseudoscalar (axion) field,  SU(N) symmetric Yang-Mills field and color aether. This theory has been applied to the model, which describes evolution of the anisotropic early Universe in that era, when two phase transitions have just been realized. The first (hypothetical) event was the spontaneous polarization of the color aether, i.e., the event that resulted in the lining up of the SU(N) symmetric vector field multiplet $U^j_{(a)}$ along the timelike unit vector field $U^j$ attributed to the canonic dynamic aether velocity four-vector \cite{3}. From the mathematical point of view, we considered the relation $U^j_{(a)}= q_{(a)}U^j$, where $q_{(a)}$ is the co-vector in the group space, and the subscript $(a)$ runs in the range $(1,..., N^2-1)$. The second phase transition, which was assumed in \cite{1,2}, could be the parallelization of the Yang-Mills potentials $A_m^{(a)}$ in the group space, i.e., $A_m^{(a)} = Q^{(a)} A_m$. This procedure makes the gauge field quasi-Abelian. In the presented paper we assume that the first phase transition has already taken place, but the second has not yet. In other words, the novelty of this work is that the gauge field is considered to be non-Abelian. Also, we consider now the SU(2) gauge group, and we study the non-Abelian SU(2) symmetric analog of the axionically induced generation of the electric field in the magnetized medium (see \cite{4,5,6} for details).

The paper is organized as follows. In Section II we present the theory formalism. In Section III we apply this formalism to the Bianchi-I spacetime model, we obtain the model evolutionary equations and formulate the preliminary results of analysis.

\section{The formalism}

We use the following total action functional:
$$
S {=} \int d^4 x \sqrt{{-}g} \left\{ \frac{1}{2\kappa}\left[R {+} 2\Lambda {+} \lambda \left(U^m_{(a)}U^{(a)}_m {-}1 \right) {+} {\cal K}^{ijmn}_{(a)(b)}\hat{D}_i U^{(a)}_{m}\hat{D}_j
U^{(b)}_n \right] + \right.
$$
\begin{equation}
\label{SUNact}
 \left. {+} \frac{1}{4} F^{(a)}_{mn} F^{mn}_{(a)} + \frac{1}{4} \sin{\phi} F^{*(a)}_{mn} F^{mn}_{(a)} {+} \frac12 \Psi^2_0  \left[V(\phi) {-} \nabla_k \phi \nabla^k \phi \right] \right\} \,.
\end{equation}
Here $R$ is the Ricci scalar, $\Lambda$ is the cosmological constant, $\kappa = 8 \pi G$ is the Einstein constant ($c=1$), $\phi$ describes the pseudoscalar (axion) field; $V(\phi)$ is the potential of the axion field.
The potential $V(\phi)$ is chosen to be of the periodic form $V(\phi) = 2 m^2_A \left(1{-}\cos{\phi} \right)$.
The constant $\Psi_0$ is reciprocal to the coupling constant of the axion-gluon interactions $g_{AG}$, i.e.,  $g_{AG}= \frac{1}{\Psi_0}$.
For the multiplet of parallel vector fields the term $(U^m_{(a)}U^{(a)}_m {-}1)$ transforms into $(U^m U_m {-}1)$ thus providing the vector $U^j$ to be timelike and unit.
The constitutive tensor ${\cal K}^{ijmn}_{(a)(b)}$ is considered to contain four coupling constants:
\begin{equation}\label{constitutive1}
{\cal K}^{ijmn}_{(a)(b)} =  G_{(a)(b)} \left[C_1 g^{ij} g^{mn} {+} C_2 g^{im}g^{jn}
{+} C_3 g^{in}g^{jm} \right] + q_{(a)}q_{(b)} C_{4} U^{i} U^{j} g^{mn} \,.
\end{equation}
The gauge covariant derivative $\hat{D}_m U^{(a)}_n$ can now be reduced to
\begin{equation}
\hat{D}_m U^{(a)}_n \equiv  q^{(a)} \nabla_m U_n + g f^{(a)}_{\ (b)(c)} A^{(b)}_m  q^{(c)}U_n
\,. \label{DU}
\end{equation}
The term $\frac{1}{4} \sin{\phi} F^{*(a)}_{mn} F^{mn}_{(a)}$ in the action functional (\ref{SUNact}) is the nonlinear periodic generalization of the term $\frac{1}{4} \phi F^{*(a)}_{mn} F^{mn}_{(a)}$ introduced by Peccei and Quinn \cite{PQ}.
The Yang-Mills field strength tensors $F^{(a)}_{mn}$ and its dual are standardly expressed as follows:
\begin{equation}
F^{(a)}_{mn} = \nabla_m
A^{(a)}_n - \nabla_n A^{(a)}_m + g f^{(a)}_{\ (b)(c)}
A^{(b)}_m A^{(c)}_n \,, \quad {}^*\! F^{ik (a)} = \frac{1}{2}\epsilon^{ikls} F^{(a)}_{ls} \,.
 \label{46Fmn}
\end{equation}
Variation with respect to $A^{(a)}_i$ gives the axionically extended Yang-Mills equations
\begin{equation}\label{Col1}
\hat{D}_k \left[F^{ik}_{(a)} + \phi F^{*ik}_{(a)}\right]=  - \frac{g^2}{\kappa}   f^{(d)}_{\ (c)(a)} U^{(c)}_k  {\cal K}^{imkn}_{(d)(b)} \hat{D}_m U^{(b)}_n \,.
\end{equation}
We keep in mind three details: first, for the SU(2) symmetry the group constants coincide with the Levi-Civita symbols, $f_{(a)(b)(c)}= \varepsilon_{(a)(b)(c)}$; second,
$\varepsilon^{(a)(b)(c)} \varepsilon_{(a)(b^{\prime})(c^{\prime})} =
\delta^b_{b^{\prime}} \delta^c_{c^{\prime}} {-} \delta^b_{c^{\prime}} \delta^c_{b^{\prime}}$; third, we work with the Landau-type gauge conditions $A^{(a)}_m U^m = 0$. With these assumptions we obtain
\begin{equation}\label{3Col4}
\nabla_k F^{ik (a)} + g f^{(a)}_{(b)(c)} A^{(b)}_k F^{ik (c)}
 = - F^{*ik}_{(a)}\nabla_k \phi  - \frac{g^2 C_1}{\kappa} A^{i(b)}\left[\delta^{(a)}_{(b)} - q^{(a)} q_{(b)} \right]   \,.
\end{equation}
For the illustration, we consider the Bianchi-I anisotropic homogeneous spacetime with the metric
\begin{equation}
ds^2 = dt^2 - a^2(t)dx^2 - b^2(t)dy^2 - c^2(t)dz^2  \,.
\label{Bianchi0}
\end{equation}
We assume that all the unknown model state functions inherit the spacetime symmetry and depend on the cosmological time only.
The global unit timelike vector $U^i$ is assumed to be of the form $U^i=\delta^i_t$. The metric (\ref{Bianchi0}) provides the covariant derivative to be symmetric, the acceleration  four-vector to be vanishing, and the expansion scalar $\Theta$ to have very simple form:
\begin{equation}
\nabla_m U_n = \nabla_n U_m = \frac12 \dot{g}_{mn}   \,, \quad  a_i \equiv U^k \nabla_k U_i =0 \,, \quad
\Theta(t) \equiv \nabla_k U^k = \frac{\dot{a}}{a} + \frac{\dot{b}}{b} +\frac{\dot{c}}{c} \,.
\label{B2}
\end{equation}

\section{SU(2) model on the Bianchi-I platform}

In the model with SU(2) symmetry we deal with the triplet of four-vectors $A^{(a)}_k$ (12 unknown functions). Using the analogs of the Landau gauge $A^{(a)}_m U^m = 0$ we reduce the number of unknown functions to nine. Also, keeping in mind that all the state functions have to depend on the cosmological time only, we are plunging into a new paradigm. In the U(1) symmetric models attributed to the electromagnetic theory we introduced a static magnetic field $B_0 = F_{12} = const$ using the nonhomogeneous potentials $A_2= \frac12 B_0 x^1$, $A_1= - \frac12 B_0 x^2$. In the SU(2) symmetric model with $A^{(a)}_j(t)$ we have principally another possibility. We consider the SU(2) symmetric analog of the magnetic field to have the structure
$F^{(a)}_{12}(t) = - g \delta^{(a)}_{(3)} A_2^{(1)}(t) A_1^{(2)}(t)$.
Assuming that the analog of the magnetic field can produce the analog of the electric field, we have to introduce the potential $A^{(3)}_3 $   and thus the electric type term
$F^{(a)}_{03} = \delta^{(a)}_{(3)} \ {\dot{A}}^{(3)}_3$. Also, we consider that $q^{(a)} = \delta^{(a)}_{(3)}$.
In other words, our ansatz is that in the model under consideration we can operate with  three independent potentials $A_2^{(1)}(t)$, $A_1^{(2)}(t)$ and $A_3^{(3)}(t)$; the corresponding six non-vanishing Yang-Mills field components are
$$
F^{(a)}_{01} = \delta^{(a)}_{(2)} \ {\dot{A}}^{(2)}_1 \,, \quad F^{(a)}_{02} = \delta^{(a)}_{(1)} \ {\dot{A}}^{(1)}_2 \,, \quad F^{(a)}_{03} = \delta^{(a)}_{(3)} \ {\dot{A}}^{(3)}_3 \,,
$$
\begin{equation}
F^{(a)}_{12} = - g \delta^{(a)}_{(3)} A_2^{(1)}A_1^{(2)} \,, \quad  F^{(a)}_{13} =  g \delta^{(a)}_{(1)} A_1^{(2)}A_3^{(3)} \,, \quad F^{(a)}_{23} = - g \delta^{(a)}_{(2)} A_2^{(1)}A_3^{(3)} \,.
\label{3x3}
\end{equation}
For this non-Abelian gauge field configuration the equations (\ref{3Col4}) gives three nontrivial equations:
$$
\frac{d}{dt}\left[\left(\frac{bc}{a}\right) \frac{d }{dt} A_1^{(2)} \right] {+} g^2 A_1^{(2)} \left[\left(\frac{c}{ab}\right)\left(A_2^{(1)} \right)^2 {+} \left(\frac{b}{ac}\right) \left(A_3^{(3)} \right)^2  {-} \frac{C_1}{\kappa} \left(\frac{bc}{a}\right) \right]  = {-} g \dot{\phi} \cos{\phi} A_2^{(1)} A_3^{(3)}  \,,
$$
$$
\frac{d}{dt}\left[\left(\frac{ac}{b}\right) \frac{d }{dt} A_2^{(1)} \right] {+} g^2 A_2^{(1)} \left[\left(\frac{c}{ab}\right)\left(A_1^{(2)} \right)^2 {+} \left(\frac{a}{bc}\right) \left(A_3^{(3)} \right)^2  {-} \frac{C_1}{\kappa} \left(\frac{ac}{b}\right) \right]  = {-} g \dot{\phi} \cos{\phi} A_1^{(2)} A_3^{(3)}  \,,
$$
\begin{equation}
\frac{d}{dt}\left[\left(\frac{ab}{c}\right) \frac{d }{dt} A_3^{(3)} \right] {+} g^2 A_3^{(3)} \left[\left(\frac{b}{ac}\right)\left(A_1^{(2)} \right)^2 {+} \left(\frac{a}{bc}\right) \left(A_2^{(1)} \right)^2  \right]  = {-} g \dot{\phi} \cos{\phi} A_1^{(2)} A_2^{(1)}  \,.
\label{3x33}
\end{equation}
The master equation for the axion field can be now reduced to
\begin{equation}
\frac{d^2 \phi}{dt^2} {+}  \frac{d}{dt} \left[ \log{(abc)} \right]\ \frac{d \phi}{dt} {+} m^2_A \sin{\phi} = \left(\frac{g \cos{\phi}}{\Psi^2_0 abc}\right) \  \frac{d}{dt} \left[A^{(2)}_1 A^{(1)}_2 A^{(3)}_3 \right] \,.
\label{3x4}
\end{equation}
The format of this note does not allow us to analyze in detail the equation for the aether velocity four-vector $U^j$. We can only confirm that there exist the solution $U^j = \delta^j_0$ to the corresponding equation (see \cite{1,2} for details).
When we deal with the gravity field equations, we obtain analogs of the equations derived and analyzed in \cite{1,2}.

\section{Outlook}

The set of four equations (\ref{3x33}), (\ref{3x4}), derived above, describes the SU(2) symmetric  model of interaction of the axion and non-Abelian gauge fields. Clearly, when the product $A_1^{(2)}A_2^{(1)}$ is nonvanishing and thus one deals with the SU(2) analog of the magnetic field $F^{(3)}_{12} \neq 0$, the axion field creates the source in the right-hand side of the equation for $A_3^{(3)}$, i.e.,  the SU(2) analog of the electric field $F^{(3)}_{03}$ inevitably appears. The profiles of the corresponding solutions will be analyzed in the next work.

\vspace{5mm}
\noindent
{\bf Acknowledgments}

\noindent
The work was supported by the Russian Science Foundation (Grant No 21-12-00130)

\end{document}